\documentclass[11pt,a4paper]{article}

\pdfoutput=1

\usepackage{jcappub}
\usepackage{graphicx}
\usepackage{float}
\usepackage{slashed}
\usepackage[font=small, labelfont=bf]{caption}
\usepackage[labelformat=simple]{subcaption}

\usepackage{amsmath, amsthm, amssymb, amsfonts}
\usepackage{multirow}
\usepackage[numbers,sort&compress]{natbib}
\usepackage{hyperref}

\newcommand{\be}{\begin{equation}}
\newcommand{\ee}{\end{equation}}
\newcommand{\bes}{\begin{equation*}}
\newcommand{\ees}{\end{equation*}}
\newcommand{\bea}{\begin{eqnarray}}
\newcommand{\eea}{\end{eqnarray}}

\newcommand{\mc}{\mathcal}
\newcommand{\vo}{\mathcal{V}}
\def\PBH{{\scriptscriptstyle \rm PBH}}
\def\CMB{{\scriptscriptstyle \rm CMB}}
\def\GW{{\scriptscriptstyle \rm GW}}
\def\LISA{{\scriptscriptstyle \rm LISA}}
\def\K3{{\scriptscriptstyle \rm K3}}

\def\W{{\scriptscriptstyle \rm W}}
\def\M{{\scriptscriptstyle \rm M}}

\title{Secondary GWs and PBHs in string inflation: formation and detectability}

\author[a,b]{Michele Cicoli,}
\author[a,b]{Francisco G. Pedro,} 
\author[a,b]{Nicola Pedron}

\affiliation[a]{\small Dipartimento di Fisica e Astronomia, Universit\`a di Bologna, \\ via Irnerio 46, 40126 Bologna, Italy}
\affiliation[b]{\small INFN, Sezione di Bologna, viale Berti Pichat 6/2, 40127 Bologna, Italy}

\emailAdd{michele.cicoli@unibo.it}
\emailAdd{francisco.soares@unibo.it}
\emailAdd{nicola.pedron2@unibo.it}

\abstract{We derive the spectrum and analyse the detectability prospects of secondary gravity waves (GWs) associated to primordial black hole (PBH) production in a class of string inflationary models called Fibre Inflation. The inflationary potential features a near inflection point that induces a period of ultra slow-roll responsible for an enhancement of the scalar perturbations which can lead to PBHs with different masses and contributions to dark matter (DM) in agreement with current observational bounds, including CMB constraints on the scalar spectral index and the tensor-to-scalar ratio. This enhancement of the curvature perturbations sources secondary GWs which can be detected by either LISA, ET or BBO, depending on the GW frequency but regardless of the amount of PBH DM since secondary GWs remain detectable even if the PBH contribution to DM is exponentially suppressed. The possibility to see a secondary GW signal is instead due to the presence of an ultra slow-roll epoch between CMB horizon exit and the end of inflation.}

\begin{document}
\maketitle

\section{Introduction}
\label{Intro}

One of the most pressing problems in modern physics concerns the nature of dark matter (DM). A possible solution is that DM is composed of primordial black holes (PBHs) \cite{Hawking:1971ei}. This is a compelling possibility since it relies neither on modifications of gravity nor on physics beyond the Standard Model (SM). With this assumption, these black holes should be much lighter and smaller than astrophysical ones, implying that their formation cannot be the result of stellar collapse. PBHs are formed during the radiation or matter eras when some large density fluctuations, produced during inflation, re-enter the horizon and undergo gravitational collapse \cite{Ivanov:1994pa}. At present the values of allowed PBH masses are very constrained \cite{Bartolo:2018evs}, if we assume that they constitute a significant fraction of the total DM abundance. There are however a couple of windows where constraints are absent and PBHs could still constitute all of the DM. One such window lies around $M_\PBH \sim 10^{-12} M_\odot$ and the other around $M_\PBH \sim 10^{-15} M_\odot$. 

A signature of the presence of PBHs might be encoded in the stochastic background of secondary gravitational waves (GWs) associated with the large curvature perturbations that lead to their formation \cite{Ananda:2006af,Baumann:2007zm,Espinosa:2018eve,Kohri:2018awv} (for a recent review see \cite{Domenech:2021ztg}). Thus current and future interferometers can test the PBH hypothesis experimentally. In particular the LISA interferometer \cite{Robson:2018ifk,Thrane:2013oya,Bavera:2021wmw} is well suited for this job since, as pointed out in \cite{Bartolo:2018evs}, a $\delta$-function power spectrum peaked at the frequency where LISA sees its maximum sensitivity would correspond to PBH masses around $10^{-12} M_\odot$. Other interferometers that might be able to detect this GW background are BBO \cite{Thrane:2013oya} and ET \cite{Bavera:2021wmw,Maggiore:2019uih} which extend to larger frequencies associated to a lower PBH mass range.

When discussing PBH formation it is also important to present a microscopic model that allows for such a process to occur. Most of the simplest single-field models are based on a near inflection point which leads to a period of ultra slow-roll (USR) responsible for the amplification of the density perturbations at momentum scales which are higher than the CMB ones \cite{Garcia-Bellido:2017mdw,Ezquiaga:2017fvi,Germani:2017bcs,Ballesteros:2017fsr}. One of the shortcomings of these constructions is in general the need to fine tune the underlying parameters in order to induce a near inflection point at the desired scales. A better approach relies instead on considering inflationary models with a consistent UV embedding which can justify the required tuning freedom and its stability against quantum corrections. In this regard, a very promising class of single-field models is Fibre Inflation \cite{Cicoli:2008gp, Broy:2015zba, Cicoli:2016chb} which arises within the framework of type IIB string compactifications. These models are characterised by three appealing features: $(i)$ their parameters are functions of background flux quanta and Calabi-Yau topological data which are tunable in the string landscape; $(ii)$ the robustness of the tuning against radiative corrections is guaranteed by the presence of an approximate rescaling symmetry \cite{Burgess:2016owb, Burgess:2014tja, Burgess:2020qsc}; $(iii)$ Calabi-Yau constructions with explicit brane setup, tadpole cancellation and moduli stabilisation provide concrete realisations of the inflationary potential \cite{Cicoli:2011it, Cicoli:2016xae, Cicoli:2017axo}. 

The typical potential of Fibre Inflation models is characterised by a plateau region which resembles Starobinsky inflation \cite{Starobinsky:1980te} and $\alpha$-attractors \cite{Kallosh:2013maa,Kallosh:2017wku}. However, as shown in \cite{Cicoli:2018asa}, its structure is rich enough to induce a near inflection point at large momentum scales. Ref. \cite{Cicoli:2018asa} presented a particular Fibre Inflation model which leads to PBH masses around $M_\PBH \sim 10^{-15} M_\odot$ which can constitute all of the DM. The main shortcoming of the model of \cite{Cicoli:2018asa} is that the scalar spectral index $n_s$ at CMB scale is more than $3\sigma$ away from Planck data \cite{Planck:2018vyg}, and this tension increases when the PBH mass is enhanced. Notice that obtaining the right observables at CMB scales and sufficient PBH production is a general challenge of single-field models, as discussed in \cite{Ballesteros:2020qam}. It is however worth pointing out that the work of \cite{Cicoli:2018asa} should be considered just as a proof of concept regarding the possibility to generate PBHs in Fibre Inflation since it considered just a simplified version of the scalar potential of the explicit Calabi-Yau examples of \cite{Cicoli:2011it, Cicoli:2016xae, Cicoli:2017axo} where some coefficients have been set to zero by hand. In this paper we shall show that a more detailed study of the inflationary potential of these models can allow for the production of PBHs which can form a considerable fraction of DM, basically at any mass scale $M_\PBH \lesssim 10^{-12} M_\odot$, and in a way compatible with $n_s$ within $1\sigma$ from the observed CMB value.

We will then analyse the associated production of secondary GWs in Fibre Inflation models finding the following interesting results which we think to be qualitatively valid more in general for any single-field model with a plateau and a near inflection point with a similar tuning freedom:
\begin{enumerate}
\item PBH DM in agreement with present observational bounds (including CMB constraints on $n_s$) can be generated at any mass scale in the range $10^{-17} M_\odot \lesssim M_\PBH \lesssim 10^{-11} M_\odot$. Moreover, in Fibre Inflation models PBH production due to USR correlates with primordial tensor modes at CMB scales which are at the edge of detectability, $r\simeq 0.01$-$0.02$, a factor of $2$ larger than the prediction for cases without USR where $r \simeq 0.007$ \cite{Cicoli:2020bao};

\item Depending on the PBH mass, the peak of secondary GWs can be at any frequency in the range $1\,{\rm mHz}\lesssim f_\GW^{\rm peak} \lesssim 1\,{\rm Hz}$, resulting in a potential detection by either LISA, ET or BBO;

\item The detectability of secondary GWs does not depend on the amount of PBH DM since it holds true even if the PBH contribution to DM is exponentially suppressed. It is instead tied to the presence of a period of USR.
\end{enumerate}

Our findings imply that if DM is made of PBHs around $M_\PBH \sim 10^{-12} M_\odot$ or $M_\PBH \sim 10^{-15} M_\odot$, then secondary GWs should be seen respectively by LISA or BBO, while a lower PBH mass (constrained by observations) would be associated to GWs detectable by ET. This remains true even if DM is not made of PBHs but the inflationary potential features a near inflection point which amplifies the density perturbations via an USR epoch. Thus a non-observation of any signal by future interferometers can, not just rule out the PBH DM hypothesis, but also constrain the form of the underlying inflationary potential.

This paper is organised as follows. In Sec. \ref{fibrinfl} we briefly review some aspects of Fibre Inflation models and present the functional form of the inflationary potential. In Sec. \ref{pbhfibrinfl} we discuss PBH formation and obtain the scalar power spectrum for the curvature perturbations, together with the associated predictions for CMB observables like the scalar spectral index and the tensor-to-scalar ratio. In Sec. \ref{sgw} we shortly recap the theory of secondary GWs, while in Sec. \ref{GWinFI} we first derive the spectrum of secondary GWs produced in Fibre Inflation and then we compare it with interferometry sensitivity curves discussing how the detection of secondary GWs depends on the fraction of PBH DM. Finally we present our conclusions in Sec. \ref{concl}.

\section{Fibre Inflation in a nutshell}
\label{fibrinfl}

Fibre Inflation is a class of models built within the framework of type IIB flux compactifications. Its name derives from the geometric properties of the extra dimensions which feature a fibration structure. The inflaton is one of the so-called K\"ahler moduli which determine the size of the dimensionless Calabi-Yau volume $\vo$ in string units. This quantity, together with the string coupling $g_s$ set by the value of the dilaton, controls the robustness of the 4D effective field theory since perturbative corrections beyond the tree-level approximation arise as either string loops weighted by positive powers of $g_s$, or as higher derivative $\alpha'$ effects proportional to inverse powers of $\vo$. Hence control over the EFT requires always to focus on values of the dilaton and the K\"ahler moduli such that $g_s\ll 1$ and $\vo \gg 1$. 

At leading order the inflaton potential vanishes due to the typical no-scale structure of type IIB constructions which induces an effective shift symmetry \cite{Burgess:2016owb, Burgess:2014tja, Burgess:2020qsc} suitable to protect the flatness of the inflationary potential generated by subdominant loop and $\alpha'$ corrections \cite{Cicoli:2008gp, Broy:2015zba, Cicoli:2016chb}. This class of inflationary models has already received a lot of attention. Previous studies revealed that Fibre Inflation can be embedded into globally consistent Calabi-Yau compactifications with moduli stabilisation \cite{Cicoli:2011it, Cicoli:2016xae, Cicoli:2017axo} where geometrical constraints can allow for an inflaton field range around $5$ Planck units \cite{Cicoli:2018tcq} corresponding to a tensor-to-scalar ratio around $r\simeq 0.007$ \cite{Cicoli:2020bao}. Even if the underlying theory is multi-field, the effective inflationary dynamics is fully single-field since fields which are heavier than the inflaton sit at their minima, while lighter modes have been shown to act just as spectators \cite{Cicoli:2019ulk, Cicoli:2021yhb, Cicoli:2021itv}. Depending on the exact duration of inflation, the pre-inflationary dynamics can be characterised by a CMB power loss at large angular scales \cite{Cicoli:2013oba, Pedro:2013pba, Cicoli:2014bja}, while the post-inflationary evolution is determined by the perturbative decay of the inflaton \cite{Cabella:2017zsa, Cicoli:2018cgu} (preheating effects have instead been shown to be negligible \cite{Antusch:2017flz}) which can successfully avoid the overproduction of axionic dark radiation \cite{Cicoli:2018cgu}, producing a SM thermal bath with a temperature below the upper bound coming from decompactification due to finite-temperature effects \cite{Anguelova:2009ht}. 

For $g_s\ll 1$ and $\vo \gg 1$ the contributions which arise at leading order in the string loop and $\alpha'$ expansions are 1-loop open string winding corrections \cite{Berg:2005ja, Berg:2007wt, Cicoli:2007xp}. In the explicit Calabi-Yau models of \cite{Cicoli:2011it, Cicoli:2016xae, Cicoli:2017axo}, these loop effects, together with an appropriate sector which uplifts the vacuum energy to positive values \cite{Kachru:2003aw, Cicoli:2015ylx, Cicoli:2012fh, Gallego:2017dvd}, generate an inflationary potential for the canonically normalised inflaton $\phi$ of the form:
\be
V_{\rm inf} = V_0 \left[ C_1 - e^{-\phi/\sqrt{3}} \left( 1 - \frac{C_2}{1 - C_3\, e^{-\phi/\sqrt{3}}} \right) 
+ e^{2 \phi/\sqrt{3}} \left( C_4 - \frac{C_5}{1 + C_6\, e^{\phi\sqrt{3}}} \right) \right],
\label{vinfl}
\ee
where $C_i$ $\forall i=1,...,6$ is a tunable coefficient in the string landscape since it depends on combinations of microscopic parameters of the compactification like Calabi-Yau intersection numbers, background and gauge fluxes, instanton wrapping numbers and ranks of condensing gauge groups.

\begin{figure}
    \centering
    \includegraphics[width=0.7\textwidth]{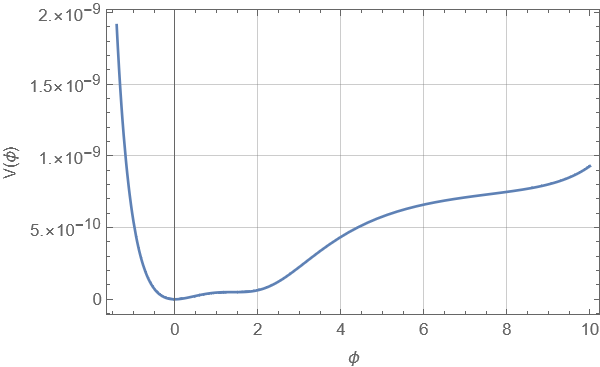}
    \caption{Inflationary potential (\ref{vinfl}) for the parameter set $\mc{P}_1$ of Tab. \ref{fib_infl_para}.}
    \label{inflpot}
\end{figure}

The potential (\ref{vinfl}) has enough tuning freedom to allow for an almost flat region at high field values, corresponding to the usual slow-roll behaviour, and a near inflection point close the minimum (parametrised by $C_5$ and $C_6$), that induces an USR dynamics. In Tab. \ref{fib_infl_para} we list the illustrative parameter sets used in the analyses performed in the following sections, and Fig. \ref{inflpot} shows the inflationary potential for the set $\mc{P}_1$.

\begin{table}
    \centering
    \begin{tabular}{c|c|c|c|c}
       & $W_0$ & $D_\W$ & $G_\W$ & $R_\W$ \\ \hline
       $\mc{P}_1$ & 14.2  & $5.90\cdot10^{-7}$ & $2.949\cdot 10^{-2}$ & $0.67381098$  \\ \hline
       $\mc{P}_2$ & 12.6  & $4.00\cdot10^{-7}$ & $3.030\cdot 10^{-2}$ & $0.69435560$  \\ \hline
       $\mc{P}_3$ & 12.1  & $3.30\cdot10^{-7}$ & $3.065\cdot 10^{-2}$ & $0.70292692$  \\ \hline
       $\mc{P}_4$ & 11.4 & $2.39\cdot10^{-7}$ & $3.104\cdot 10^{-2}$ & $0.71296774$  \\ \hline
       $\mc{P}_5$ & 13.7  & $5.20\cdot10^{-7}$ & $2.938\cdot 10^{-2}$ & $0.67101003$  \\ \hline
       $\mc{P}_6$ & 14.9  & $6.58\cdot10^{-7}$ & $2.830\cdot 10^{-2}$ & $0.64335706$  \\
    \end{tabular}
    \caption{Different parameter sets for the inflationary potential (\ref{vinfl}). Following the notation of \cite{Cicoli:2018asa}, in all cases $C_2=0.5$, $C_3 = 1/\sqrt{\lambda}$, $C_4 = 25 \gamma D_\W$, $C_5 = 25 \gamma G_\W$ and $C_6 = \gamma R_\W$ with $\gamma \equiv \lambda^{3/2}/\vo$ for $\lambda= 14.3$ and $\vo=1000$. $C_1$ is instead found by requiring $V_{\rm inf}=0$ at the minimum and $V_0 \equiv W_0^2/\left(25 \gamma^{1/3}\vo^{10/3}\right)$.}
    \label{fib_infl_para}
\end{table}

\section{PBHs in Fibre Inflation and CMB observables}
\label{pbhfibrinfl}

As already pointed out, PBH formation is possible if the inflaton undergoes a period of USR \cite{Motohashi:2017kbs,Inomata:2017okj} caused by the presence of a near inflection point in the potential, with a resulting enhancement of the scalar power spectrum at PBH formation scales. Consequently some large and relatively rare density perturbations undergo a gravitational collapse at horizon re-entry, during the radiation or matter phases, to form PBHs. In this work we will assume PBH formation in the radiation era. The value of the enhancement of the scalar power spectrum at PBH formation scales can be inferred considering the fraction of the total energy density in PBHs with mass $M_\PBH$ at PBH formation time which takes the form (assuming a Gaussian distribution for the curvature perturbations):
\be
\beta (M_\PBH) \equiv \frac{\rho_\PBH (M_\PBH)}{\rho_{\rm tot}} = \frac{1}{\sqrt{2\pi} \sigma_\M} \int_{\zeta_c}^{\infty} \mathrm{d}\zeta\, e^{-\frac{\zeta^2}{2\sigma_\M^2}} \approx \frac{\sigma_\M}{\sqrt{2\pi} \zeta_c}\, e^{-\frac{\zeta_c^2}{2\sigma_\M^2}}\,,
\ee
where $\zeta_c$ is the critical value for the gravitational collapse to take place, and we have used the approximation $\sigma_\M \ll \zeta_c$\footnote{For non-Gaussian corrections to PBH formation and the induced GW spectrum see \cite{Atal:2018neu,Atal:2021jyo}. Non-Gaussianities will induce a slight change of $\zeta_c$ with respect to the Gaussian estimate, which can easily be accommodated in the present model.}. Moreover if we redshift and rescale up to present time we obtain the relation:
\be
\beta (M) \approx 10^{-8} \sqrt{\frac{M_\PBH}{M_\odot}}\, \mathfrak{f}_\PBH (M_\PBH)\ ,
\ee
where $\mathfrak{f}_\PBH (M_\PBH)$ is the fraction of PBHs with mass $M_\PBH$ contributing to the total DM abundance. Then putting these two results together and setting $\zeta_c = 0.6$ \cite{Atal:2019erb} we obtain: 
\be
\frac{\sigma_\M}{\sqrt{2\pi} \cdot 0.6}\, e^{-\frac{0.36}{2\sigma_\M^2}} \approx 10^{-8} \sqrt{\frac{M_\PBH}{M_\odot}}\, \mathfrak{f}_\PBH (M_\PBH)\ .
\label{sigma}
\ee
If we assume a distribution peaked at $M_\PBH = 10^{-12} M_\odot$, and that $\mathfrak{f}_\PBH (10^{-12} M_\odot) = 1$ (i.e. all of DM is made of PBHs of this mass), then we obtain $\sigma_\M \simeq 0.078$. Therefore since $\sigma_\M^2$ is related to the two-point function, $\sigma_\M^2 \sim \langle \zeta \zeta \rangle$, this means that the peak in the power spectrum has to be of order $\mc{O}(10^{-3})$ at PBH formation scales, a value which is $6$ orders of magnitude higher than that at CMB scales. Notice that this remains true even for $\mathfrak{f}_\PBH \ll 1$ due to the exponential dependence on $\sigma_\M$ in (\ref{sigma}) (for example $\sigma_\M \simeq 0.059$ for $\mathfrak{f}_\PBH (10^{-12} M_\odot) \simeq 10^{-10}$). The possibility of having such an enhancement arises due to the tuning freedom of the inflationary potential that allows to have a near inflection point after the point in field space corresponding to CMB horizon exit. 

The value of the PBH mass is related to the number of efoldings that elapsed between the moments of CMB and PBH horizon exit via \cite{Motohashi:2017kbs}:
\be
\Delta N_\CMB^\PBH = \ln \left( \frac{a_\PBH H_{\rm inf}}{a_\CMB H_{\rm inf}} \right) = 18.4 - \frac{1}{12} \ln \left( \frac{g_{*f}}{g_{*0}} \right) 
+ \frac12 \ln \gamma_c - \frac12 \ln \left( \frac{M_\PBH}{M_\odot} \right),
\ee
where we have assumed $H$ to be constant during inflation (and equal to $H_{\rm inf}$), $g_{*f,0}$ are the number of relativistic degrees of freedom at PBH formation time and today, while $\gamma_c$ is a correction factor related to the details and the efficiency of the gravitational collapse. Assuming $\gamma_c=0.8$ (as in the case for a narrow power spectrum \cite{Germani:2018jgr}) and that the only degrees of freedom are those of the SM, so that $g_{*f} = 106.75$ and $g_{*0} = 3.36$, for PBHs of mass $10^{-12} M_\odot$, we find that PBH scales leave the horizon approximately $32$ efoldings after CMB scales, or equivalently $21$ efoldings before the end of inflation since recent studies of reheating after the end of Fibre Inflation showed that CMB scales exit the horizon $53$ efoldings before the end of inflation \cite{Cicoli:2018cgu}.

In order to compute the scalar power spectrum we need to know the evolution of the rescaled curvature perturbations, which is governed by the Mukhanov-Sasaki (MS) equation:
\be
u''_k (\tau) + \left(k^2 - \frac{z''}{z} \right) u_k(\tau) = 0\,,
\label{MS}
\ee
where the prime denotes a derivative with respect to conformal time $\tau$, $z = \sqrt{2\epsilon}\, a$ and $u_k = z\zeta_k$ is the MS variable. The Hubble slow-roll parameters:
\be
\epsilon = -\frac{\dot{H}}{H^2}\,, \qquad \eta = \frac{\dot{\epsilon}}{\epsilon H}\,, \qquad \kappa = \frac{\dot{\eta}}{\eta H}\,,
\ee
are computed solving the inflaton equation of motion, with the potential (\ref{vinfl}), coupled to the Friedmann equation and enter the expression for the MS equation as:
\be
\frac{z''}{z} = (aH)^2 \left[ 2 - \epsilon + \frac{3\eta}{2} - \frac{\epsilon \eta}{2} + \frac{\eta^2}{4} + \frac{\eta \kappa}{2} \right].
\label{zz}
\ee
The dimensionless power spectrum $\mc{P}_\zeta (k) = \frac{k^3}{2\pi^2} |\frac{u_k}{z}|^2$ is then obtained solving the MS equation numerically since the familiar slow-roll approximation:
\be
\mc{P}_\zeta^{\rm (sr)} (k) = \frac{H^2}{8\pi^2\epsilon} \bigg |_{k = aH}\,,
\label{pksr}
\ee
ceases to be valid due to super-horizon evolution during USR \cite{Germani:2017bcs,Motohashi:2017kbs}. To see this in more detail, let us consider perturbations evolving in a background with constant $\eta$ as in the USR case. We can then integrate the equation for the first slow-roll parameter $\epsilon$:
\be
\eta = \frac{\dot{\epsilon}}{\epsilon H} = \frac{\epsilon_N}{\epsilon} \qquad \Longrightarrow \qquad \epsilon (N) = \epsilon_0\, e^{\eta N}\,,
\ee
where the subscript $N$ denotes a derivative with respect to the number of efoldings $N$ such that $\mathrm{d}N = H \mathrm{d}t = aH \mathrm{d}\tau$. Then for $\eta < 0$ we have an exponential decrease of $\epsilon$. For constant $\eta = -\mc{O}(1)$ we can set $\kappa = 0$ and neglect in (\ref{zz}) the two terms proportional to $\epsilon$, obtaining the following approximation:
\be
\frac{z''}{z} \approx (aH)^2 \left [ 2 + \frac32 \eta + \frac{\eta^2}{4} \right]\,.
\ee
We can then recast the MS equation as a function of $N$ as:
\be
u_{NN} + (1-\epsilon) u_N + \left [ \left ( \frac{k}{aH} \right )^2 + \left ( 2 + \frac32 \eta + \frac{\eta^2}{4} \right ) \right ] u = 0\,.
\ee
For super-horizon scales $k \ll aH$, and assuming $\epsilon \ll 1$, this can be approximated as:
\be
u_{NN} + u_N - \left ( 2 + \frac32 \eta + \frac{\eta^2}{4} \right )u = 0\,.
\label{ueq}
\ee
We look for solutions of the form $u(N) = u_0\, e^{\lambda N}$, with $\lambda$ to be determined and $u_0$ an integration constant. Inserting this Ansatz into (\ref{ueq}) we get:
\be
\lambda^2 + \lambda - \left( 2 + \frac32 \eta + \frac{\eta^2}{4} \right) = 0 \qquad \Leftrightarrow \qquad \lambda_\pm = \frac{-1 \pm |\eta + 3|}{2}\,.
\ee
We now turn to the physical variable related to the curvature perturbations, i.e. the scalar $\zeta = u/z$. Using the definition for the MS variable we get:
\be
z = \sqrt{2\epsilon}\,a \propto e^{N \left( 1 + \frac{\eta}{2} \right)} \qquad \Longrightarrow \qquad \zeta(N) \propto e^{N \omega_\pm (\eta)}\,,
\ee
where:
\be
\omega_\pm (\eta) \equiv \lambda_\pm - 1 - \frac{\eta}{2} = \frac{\pm |\eta + 3| - (\eta + 3)}{2}\,.
\ee
Therefore the behaviour of the curvature perturbations is dictated by the sign of the expression $\omega_\pm (\eta)$ in the exponent. The most general solution is then given by:
\be
\zeta (N) = \zeta_0\, e^{\omega_+ N} + \overline{\zeta}_0\, e^{\omega_- N}\,,
\ee
where $\zeta_0$ and $\overline{\zeta}_0$ are integration constants. We see that $\omega_+$ describes a constant mode for $\eta \geq -3$ and a growing mode for $\eta < -3$, while $\omega_-$ describes a decaying mode for $\eta \geq -3$ and a constant mode for $\eta < -3$. Therefore in an USR epoch characterised by $\eta < -3$, the curvature perturbations grow exponentially and the slow-roll approximation (\ref{pksr}) cannot be trusted since it assumes that there is no super-horizon evolution.

\begin{figure}
    \centering
    \includegraphics[width=0.9\textwidth]{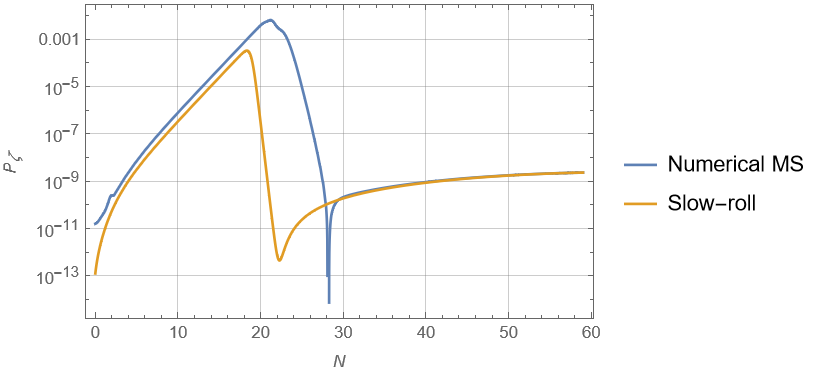}
    \caption{Scalar power spectrum for the parameter set $\mc{P}_1$ of Tab. \ref{fib_infl_para} as a function of the number of efoldings before the end of inflation. The blue line represents the numerical solution of the MS equation (\ref{MS}) while the orange line is the slow-roll approximation (\ref{pksr}).}
    \label{pps}
\end{figure}

The scalar power spectrum for parameter set $\mc{P}_1$ of Tab. \ref{fib_infl_para} is shown in Fig. \ref{pps}, together with the slow-roll approximation (\ref{pksr}). The peak is at scales leaving the horizon about $21$ efoldings before the end of inflation, corresponding to PBHs with mass $M_\PBH \simeq 10^{-12} M_\odot$, and it is of order $\mc{O}(10^{-3})$ as previously discussed, such that PBHs can account for all DM.

Similar results have already been obtained in \cite{Cicoli:2018asa} for $M_\PBH \simeq 10^{-15} M_\odot$ with however a more than $3\sigma$ tension with the observed value of the scalar spectral index $n_s$ which gets worse when PBH scales get closer to CMB scales, as for the case corresponding to $M_\PBH \simeq 10^{-12} M_\odot$. We realised that this observational problem was just a consequence of setting in the inflationary potential (\ref{vinfl}) the coefficient $C_4=0$ by hand in \cite{Cicoli:2018asa}, a choice taken for simplicity of presentation given that the aim of that paper was just to show that Fibre Inflation models can lead to enough PBH formation. As can be seen from Tab. \ref{fib_infl_para}, in this paper we have instead allowed a non-zero $C_4$ whose tuning (guaranteed by the underlying flux landscape) can lead to a scalar spectral index $n_s = 1 + \frac{d\ln \mc{P}_k}{d\ln k} = 0.965$, which is fully compatible with observations within $1\sigma$ \cite{Planck:2018vyg}. To compute it we did not rely on the slow-roll approximation due to the super-horizon evolution triggered by the USR phase. Hence we performed a series of linear regressions of a dense set of points in the range $N \in [52.5,53.5]$ and computed the mean value of the derivatives. The tensor-to-scalar ratio turns out to be rather large, $r \simeq 0.024$, in accordance with a trans-Planckian inflaton range \cite{Lyth:1996im}, and still within current bounds. Notice that this prediction for $r$ is about a factor of $2$ larger than the one of Fibre Inflation models which do not feature a period of USR \cite{Cicoli:2020bao}. The peak of the scalar power spectrum, the PBH mass and DM fraction, together with the associated CMB predictions, are listed in Tab. \ref{fib_infl_CMB} for all the parameter sets of Tab. \ref{fib_infl_para}. The predictions are evaluated at $53$ efoldings before the end of inflation and each parameter set reproduces the observed amplitude of the density perturbations at CMB scales.

\begin{table}
\centering
  \begin{tabular}{c|c|c|c|c|c}
   & $\mc{P}_\zeta^{\rm peak}$ & $M_\PBH \ [M_\odot]$ & $\mathfrak{f}_\PBH$ & $n_s$ & $r$ \\ \hline
   $\mc{P}_1$ & $6.6 \cdot 10^{-3}$ & $\mc{O} (10^{-12})$ & $\mc{O} (1)$ & $0.965$ & $0.024$ \\ \hline
   $\mc{P}_2$ & $6.4 \cdot 10^{-3}$ & $\mc{O} (10^{-15})$ & $\mc{O} (1)$ & $0.967$ & $0.020$ \\ \hline
   $\mc{P}_3$ & $6.2 \cdot 10^{-3}$ & $\mc{O} (10^{-16})$ & $\mc{O} (1)$ & $0.965$ & $0.018$ \\ \hline
   $\mc{P}_4$ & $6.1 \cdot 10^{-3}$ & $\mc{O} (10^{-17})$ & $\mc{O} (10^{-3})$ & $0.963$ & $0.016$ \\ \hline
   $\mc{P}_5$ & $4.1 \cdot 10^{-3}$ & $\mc{O} (10^{-12})$ & $\mc{O} (10^{-8})$ & $0.966$ & $0.022$  \\ \hline
   $\mc{P}_6$ & $1.1 \cdot 10^{-4}$ & $\mc{O} (10^{-12})$ & $0$ & $0.966$ & $0.026$ \\ 
  \end{tabular}
  \caption{Peak values of the curvature power spectrum, PBH mass and DM fraction, associated predictions for the scalar spectral index and tensor-to-scalar-ratio for the parameter sets of Tab. \ref{fib_infl_para}.}
  \label{fib_infl_CMB}
\end{table}

\section{Secondary GWs from PBHs: a brief review}
\label{sgw}

The amplification of the scalar perturbations during inflation, as needed for PBH formation, generates a stochastic background of GWs at second order in perturbation theory. More detailed discussions about the derivation of the GW energy density can be found in \cite{Ananda:2006af,Baumann:2007zm,Espinosa:2018eve,Kohri:2018awv}. In order to compute its corresponding tensor power spectrum we start from a perturbed flat FRW metric in conformal gauge:
\be
\mathrm{d}s^2 = a(\tau)^2 \left[ -(1 + 2\Psi)\,\mathrm{d}\tau^2 + \left( (1 - 2\Psi)\,\delta_{ij} + \frac12\, h_{ij} \right) \mathrm{d}x^i \mathrm{d}x^j \right],
\ee
where $\Psi$ is a Bardeen potential, $h_{ij}$ is the traceless and transverse tensor perturbation related to GWs and we ignored vector perturbations. Inserting this metric into Einstein equations and expanding up to second order, we obtain \cite{Acquaviva:2002ud}:
\be
h''_{ij} + 2\mc{H}\, h'_{ij} - \nabla^2 h_{ij} = -4 \mc{P}^{k\ell}_{ij} \mc{S}_{k\ell} \,,
\ee
where $\mc{H} = a'/a$ is the conformal Hubble parameter and $\mc{P}^{k\ell}_{ij}$ projects onto the transverse and traceless part of the source term $\mc{S}_{k\ell}$, in turn given by:
\be
\mc{S}_{k\ell} = 4 \Psi \partial_k \partial_\ell \Psi + 2 \partial_k \Psi \partial_\ell \Psi 
- \partial_k \left( \frac{\Psi'}{\mc{H}} + \Psi \right) \partial_\ell \left( \frac{\Psi'}{\mc{H}} + \Psi \right).
\ee
We can also write the scalar perturbations using a transfer function for a radiation dominated era and relate them to the initial gauge invariant curvature perturbations $\zeta_k$ as:
\be
\hat{\Psi} (\tau, \mathbf{k}) = \frac23\, T(k\tau) \zeta_k \qquad \text{with}\qquad T(x) = \frac{9}{x^2} \left [\frac{\sin \left (x/\sqrt{3} \right )}{x/\sqrt{3}} - \cos \left (x/\sqrt{3} \right ) \right],
\ee
where the hat denotes the Fourier modes of the Bardeen potential. Defining the tensor power spectrum $\mc{P}_h (k)$ as:
\be
\langle h_\mathbf{k}^{(m)} (\tau) h_\mathbf{p}^{(n)} (\tau) \rangle = \delta^{(3)} (\mathbf{k} + \mathbf{p}) \delta^{(m)(n)} \frac{2\pi^2}{k^3} \mc{P}_h (\tau,k)\ ,
\ee
where $(m)$ and $(n)$ stand for either one of two tensor polarisation states, $(+)$ or $(\times)$, its relation with the GW energy density $\Omega_\GW (\tau, k)$ is \cite{Maggiore:1999vm}:
\be
\Omega_\GW(\tau, k) \equiv \frac{\rho_\GW (\tau,k)}{\rho_c} = \frac{1}{24} \left( \frac{k}{\mc{H} (\tau) } \right)^2 \overline{\mc{P}_h(\tau,k)}\,,
\ee
where the bar denotes an average over time. Then a straightforward calculation (we refer to \cite{Ananda:2006af,Baumann:2007zm,Espinosa:2018eve,Kohri:2018awv} for all the details), after integration of Einstein equations with second order source term in the scalars, shows that the energy density of these secondary GWs at present time $\tau = \tau_0$ is given by:
\bea
\Omega_\GW (\tau_0, k) &=& \frac{c_g \Omega_{r,0}}{36} \int_{\frac{1}{\sqrt{3}}}^\infty \mathrm{d}s \int_0^{\frac{1}{\sqrt{3}}} \mathrm{d}d 
\left[ \frac{(s^2 - 1/3)(d^2 - 1/3)}{s^2 - d^2} \right]^2 \nonumber \\
&& \mc{P}_\zeta \left ( \frac{k\sqrt{3}}{2} (s+d) \right ) \mc{P}_\zeta \left( \frac{k\sqrt{3}}{2} (s-d) \right) 
\left[ \mc{I}^2_c(s,d) + \mc{I}^2_s(s,d) \right],
\label{omegagw}
\eea
where $c_g \simeq 0.4$ is a function of the relativistic degrees of freedom (assumed to be only SM ones), $\Omega_{r,0} \simeq 8 \cdot 10^{-5}$ is the radiation energy density at present time and the functions $\mc{I}_{c,s}(s,d)$ come from the analytic integration of the transfer function and read:
\bea
\mc{I}_c(s,d) &=& -36 \pi \frac{(s^2 + d^2 - 2)^2}{(s^2 - d^2)^3} \theta (s-1)\,, \\
\mc{I}_s(s,d) &=& -36 \frac{s^2 + d^2 - 2}{(s^2 - d^2)^2} \left[ \frac{s^2 + d^2 - 2}{s^2 - d^2} \ln \left( \frac{1 - d^2}{|s^2 - 1|} \right) + 2 \right],
\eea
where $\theta$ is the Heaviside step function. 

\section{Secondary GWs in Fibre Inflation: predictions and detectability}
\label{GWinFI}

In this section we analyse in detail the predictions and prospects for detectability of secondary GWs produced in Fibre Inflation in association with PBH formation. We shall split our discussion in two cases, depending on whether PBHs contribute to DM or not.

\subsection{Secondary GWs with PBHs}

The parameter sets $\mc{P}_i$ in Tab. \ref{fib_infl_para} with $i=1,2,3$ give rise to PBHs which are $100\%$ of DM in agreement with present observational bounds and for different ranges of $M_\PBH$, from $10^{-12}\,M_\odot$ to $10^{-16}\,M_\odot$. Moreover, the case $\mc{P}_4$ can also produce the largest contribution to DM allowed by Hawking evaporation, $\mathfrak{f}_\PBH \sim \mc{O}(10^{-3})$, for $M_\PBH \simeq 10^{-17}\,M_\odot$. As can be seen from Tab. \ref{fib_infl_CMB}, in all these cases the peak value of the scalar power spectrum is of order $\mc{P}^{\rm peak}_\zeta \sim \mc{O}(10^{-3})$. Using (\ref{omegagw}) to make an illustrative estimate, this gives a peak value of the GW energy density of order $\Omega^{\rm peak}_\GW h^2 \sim \mc{O}(10^{-10})$ which is large enough to be detectable by future interferometers like LISA, BBO or ET which focus on different frequencies (LISA is around $1$-$10$ mHz, BBO is around $0.1$-$1$ Hz while ET is around $1$-$100$). Notice that PBH masses in the range $10^{-12}\,M_\odot\lesssim M_\PBH\lesssim 10^{-17}\,M_\odot$ correspond exactly to secondary GWs peaked at frequencies $1\,{\rm mHz}\lesssim f_\GW^{\rm peak} \lesssim 1\,{\rm Hz}$, with higher frequencies correlated with smaller PBH masses. 

\begin{table}
\centering
  \begin{tabular}{c|c|c|c|c|c}
   & $\mc{P}_\zeta^{\rm peak}$ & $M_\PBH \ [M_\odot]$ & $\mathfrak{f}_\PBH$ & $f_\GW^{\rm peak}$ [Hz] & $\Omega_\GW^{\rm peak} h^2$ \\ \hline
   $\mc{P}_1$ & $6.6 \cdot 10^{-3}$ & $\mc{O} (10^{-12})$ & $\mc{O} (1)$ & $5.96 \cdot 10^{-3}$ & $1.08 \cdot 10^{-10}$ \\ \hline
   $\mc{P}_2$ & $6.4 \cdot 10^{-3}$ & $\mc{O} (10^{-15})$ & $\mc{O} (1)$ & $0.178$ & $9.32 \cdot 10^{-11}$ \\ \hline
   $\mc{P}_3$ & $6.2 \cdot 10^{-3}$ & $\mc{O} (10^{-16})$ & $\mc{O} (1)$ & $0.567$ & $8.99 \cdot 10^{-11}$ \\ \hline
   $\mc{P}_4$ & $6.1 \cdot 10^{-3}$ & $\mc{O} (10^{-17})$ & $\mc{O} (10^{-3})$ & $1.73$ & $8.80 \cdot 10^{-11}$  \\ 
  \end{tabular}
  \caption{Peak values of the curvature power spectrum, PBH mass and DM fraction, associated predictions for secondary GW frequency and energy density at the peak for the parameter sets $\mc{P}_i$, $i=1,..4$, of Tab. \ref{fib_infl_para}.}
  \label{fib_infl}
\end{table}

\begin{figure}[t]
    \centering
    \includegraphics[width=\textwidth]{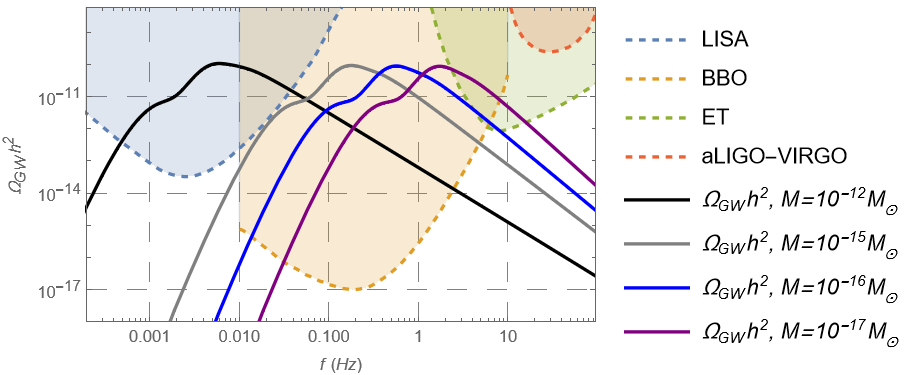}
    \caption{GW energy density as a function of frequency compared to the sensitivity curves of different forthcoming observations for the parameter sets $\mc{P}_i$, $i=1,..4$, of Tab. \ref{fib_infl_para}.}
    \label{omegaf}
\end{figure}

For each of the parameter sets $\mc{P}_i$ of Tab. \ref{fib_infl_para} with $i=1,...,4$, we performed a detailed study of the associated secondary GW signal by first solving numerically the MS equation and then inserting the scalar power spectrum in (\ref{omegagw}). In Tab. \ref{fib_infl} we report the exact values of the secondary GW frequency and energy density at the peak, while Fig. \ref{omegaf} shows the resulting GW energy density as a function of the frequency, together with the sensitivity curves of current and future interferometers \cite{Robson:2018ifk,Thrane:2013oya,Maggiore:2019uih,Bavera:2021wmw}. As anticipated by our estimate, the GW signal turns out to be within reach of LISA, BBO and ET, but not by aLIGO-VIRGO which will be sensitive to $\Omega_\GW h^2 \gtrsim 10^{-10}$ and will operate at higher frequencies. Nevertheless, this result is interesting since it implies that the secondary GW signal associated to PBH production will be tested directly in the near future. 

\subsection{Secondary GWs without PBHs}

Up to this point we focused on secondary GWs in cases where a considerable fraction of DM is made of PBHs. This assumption can however be significantly relaxed. In fact, observable secondary GWs can be obtained even when PBHs are a negligible component of the DM abundance. As already noticed in \cite{Bartolo:2018rku}, this is due to the exponential dependence of $\mathfrak{f}_\PBH$ on $\sigma_\M$ in \eqref{sigma} which implies that even a small decrease of $\sigma_\M$ makes the PBH contribution to DM exponentially suppressed. On the other hand, since $\Omega_\GW h^2 \sim \mc{P}_\zeta^2 \sim \sigma_\M^4$, such a small decrease in the amplitude of the scalar power spectrum can still lead to a sizable $\Omega_\GW h^2$. It is therefore interesting to determine the value of $\sigma_\M$, and the corresponding value of $\mathfrak{f}_\PBH$ for a given $M_\PBH$, that would still give observable secondary GWs, adjusting the magnitude of the peak of $\Omega_\GW h^2$ in such a way that it is still within the sensitivity curve of future interferometers. 

In what follows we will focus on LISA and, to quantify this statement, we shall discuss two examples, the parameter sets $\mc{P}_5$ and $\mc{P}_6$ of Tab. \ref{fib_infl_para}, which give rise to $M_\PBH\simeq 10^{-12}\,M_\odot$, corresponding to GW frequencies in the range $1$-$10$ mHz probed by LISA.

\begin{figure}
\centering
\begin{minipage}{.5\textwidth}
  \centering
  \includegraphics[width=0.9\linewidth]{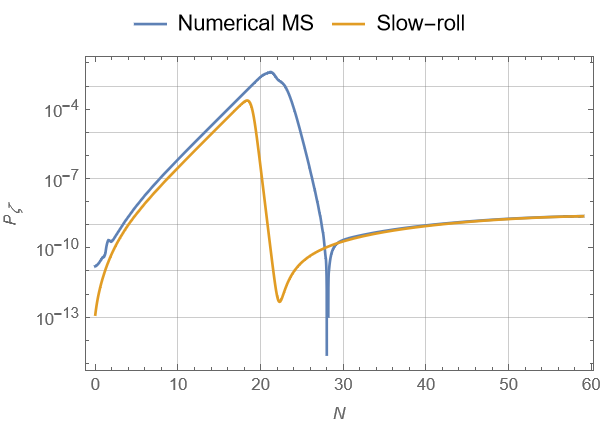}
\end{minipage}%
\begin{minipage}{.5\textwidth}
  \centering
  \includegraphics[width=0.9\linewidth]{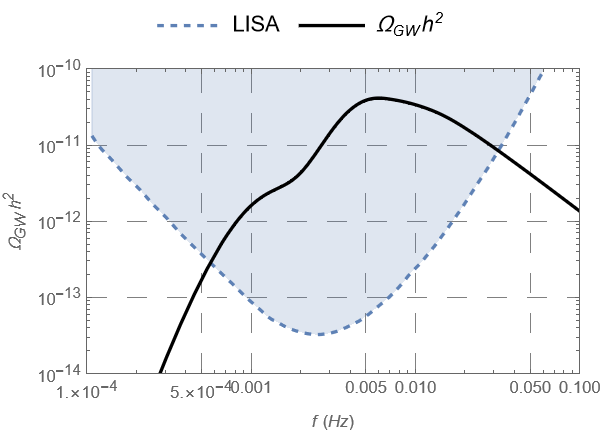}
\end{minipage}
\caption{Scalar power spectrum as a function of the number of efoldings before the end of inflation and GW energy density for the parameter set $\mc{P}_5$ of Tab. \ref{fib_infl_para}.}
\label{pps_detune}
\end{figure}

\begin{itemize}
\item \textbf{Parameter set $\mc{P}_5$}: This case leads to a PBH mass scale similar to the parameter set $\mc{P}_1$ but with $\mathfrak{f}_\PBH\ll 1$. To estimate the order of magnitude of $\mathfrak{f}_\PBH$, notice that $\mc{P}_1$ is characterised by $\mathfrak{f}_\PBH\sim\mc{O}(1)$ and $\Omega_\GW h^2 (f^{\rm min}_\LISA) = 3.58 \cdot 10^{-10}$ where $f^{\rm min}_\LISA \simeq 2.52$ mHz is the frequency of the minimum of the LISA sensitivity curve. On the other hand, $\mc{P}_5$ features $\Omega_\GW h^2 (f^{\rm min}_\LISA)= 8.24 \cdot 10^{-12}$. Recalling that $\Omega_\GW h^2 \sim \sigma_\M^4$, this implies a reduction of $\sigma_\M$ of order $\sigma_\M \rightarrow \sigma_\M/1.46$, which, when plugged into (\ref{sigma}), would give $\mathfrak{f}_\PBH \sim \mc{O}(10^{-15})$ for $M_\PBH\simeq 10^{-12} M_\odot$. 

Let us stress that this value of $\mathfrak{f}_\PBH$ is underestimated since the previous discussion concerned only the peak values of $\Omega_\GW h^2$ and $\mc{P}_\zeta$, ignoring that in (\ref{omegagw}) there is a double integration of the power spectrum that conspires to increase the value of $\Omega_\GW h^2$ from contributions at all scales. This does not make a big difference in the magnitude of the GW signal but it has a strong impact on $\mathfrak{f}_\PBH$ due to the exponential behaviour in (\ref{sigma}). Nevertheless our estimate still catches the main point: secondary GWs would still be observable even with a tiny fraction of PBH DM.

\begin{table}
  \centering
  \begin{tabular}{c|c|c|c|c|c}
   & $\mc{P}_\zeta^{\rm peak}$ & $M_\PBH \ [M_\odot]$ & $\mathfrak{f}_\PBH$ & $f_\GW^{\rm peak}$ [Hz] & $\Omega_\GW^{\rm peak} h^2$ \\ \hline
     $\mc{P}_5$ & $4.1 \cdot 10^{-3}$ & $\mc{O} (10^{-12})$ & $\mc{O} (10^{-8})$ & $5.97 \cdot 10^{-3}$ & $8.80 \cdot 10^{-11}$ \\ \hline
     $\mc{P}_6$ & $1.1 \cdot 10^{-4}$ & $\mc{O} (10^{-12})$ & $0$ & $2.80 \cdot 10^{-3}$ & $3.35 \cdot 10^{-14}$ \\ 
  \end{tabular}
	  \caption{Peak values of the curvature power spectrum, PBH mass and DM fraction, associated predictions for secondary GW frequency and energy density at the peak for the parameter sets $\mc{P}_5$ and $\mc{P}_6$ of Tab. \ref{fib_infl_para}.}
  \label{fpeak}
\end{table}

In Fig. \ref{pps_detune} we present the exact numerical results for the scalar power spectrum and the GW energy density for the parameter set $\mc{P}_5$ of Tab. \ref{fib_infl_para}. The values of the GW frequency and energy density at the peak are given in Tab. \ref{fpeak}. This case leads to $M_\PBH\sim \mc{O} (10^{-12})\,M_\odot$ and $\mathfrak{f}_\PBH \sim \mc{O} (10^{-8})$ which has been computed using the peak value of the scalar power spectrum in (\ref{sigma}). 

\newpage

\item \textbf{Parameter set $\mc{P}_6$}: This case corresponds instead to the limit where the peak of the GW energy density coincides with the minimum of the LISA curve, i.e.:
\be
\Omega^{\rm peak}_\GW h^2= \Omega^{\rm min}_\LISA h^2\,.
\ee
Fig. \ref{pps_limit} shows the numerical results for the scalar power spectrum and the GW energy density for this case (parameter set $\mc{P}_6$ of Tab. \ref{fib_infl_para}) and Tab. \ref{fpeak} lists the values of the GW frequency and energy density at the peak which matches $\Omega^{\rm min}_\LISA h^2$ within a few percent. This case features $M_\PBH\sim \mc{O} (10^{-12})\,M_\odot$ and a negligibly small PBH contribution to DM, which we estimate to be around $\mathfrak{f}_\PBH \lesssim \mathcal{O}(10^{-700})\sim 0$.

\begin{figure}
\centering
\begin{minipage}{.5\textwidth}
  \centering
  \includegraphics[width=0.9\linewidth]{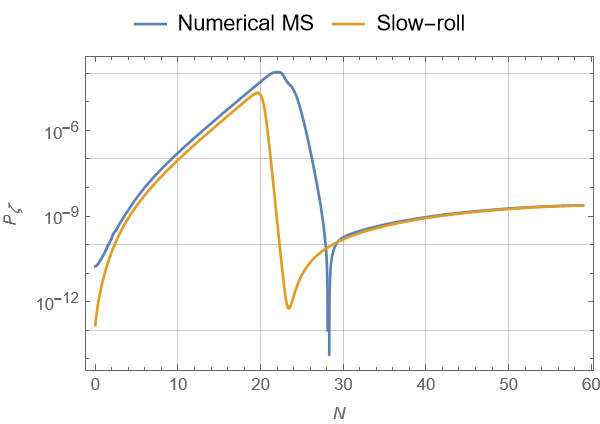}
\end{minipage}%
\begin{minipage}{.5\textwidth}
  \centering
  \includegraphics[width=0.9\linewidth]{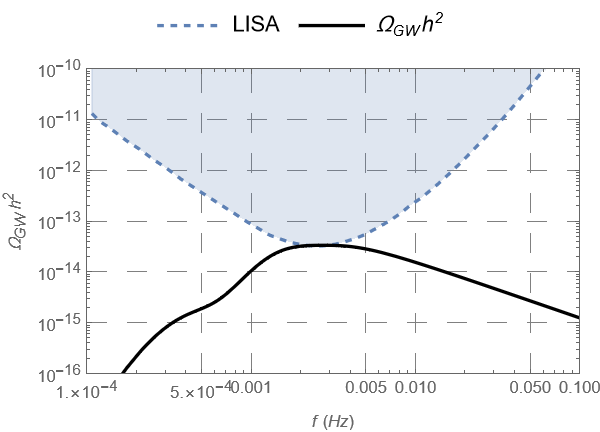}
\end{minipage}
\caption{Scalar power spectrum as a function of the number of efoldings before the end of inflation and GW energy density for the parameter set $\mc{P}_6$ of Tab. \ref{fib_infl_para}.}
\label{pps_limit}
\end{figure}
\end{itemize}

The bottom line of our analysis is therefore that, while insisting on PBH DM implies the production of secondary GWs, as stressed in \cite{Bartolo:2018evs}, the converse is not necessarily true since secondary GWs can be detectable even without PBHs contributing significantly to DM. A detectable stochastic background of GWs is instead a signature of a prolonged period of USR dynamics during the last phases of inflation, regardless of the PBH contribution to DM. It is worth stressing that, due to these general considerations, we expect this result to hold not only for Fibre Inflation models, but also for all single-field models exhibiting USR dynamics induced by a near inflection point in the inflationary potential. 

Hence a non-detection of GWs by LISA would rule out, not just the PBH DM hypothesis, but also any USR evolution during inflation on scales which cannot be probed directly by the CMB, regardless of the nature of DM. In particular, a non-detection of the stochastic background of GWs by LISA would imply a constraint on the maximum value of the power spectrum at scales leaving the horizon around $20$-$24$ efoldings before the end of inflation. Larger momentum scales could be constrained by BBO and ET, where the corresponding modes leave the horizon about $14$-$19$ efoldings before the end of inflation, as for the parameter sets $\mc{P}_i$, $i=2,3,4$ of Tab. \ref{fib_infl_para}. As already stressed, given that $\mathfrak{f}_\PBH$ is almost irrelevant when computing $\Omega_\GW h^2$, it is possible to relax the assumption that we need a PBH mass that is not constrained when we compare with GW detections. Thus the whole region $M_\PBH \in [10^{-17}, 10^3] \ M_\odot$ becomes available and having a peak in the curvature power spectrum at the corresponding scales would still give a significant GW signal. These masses correspond to scales leaving the horizon from $N = 15.3$ to $N = 38.3$ efoldings before the end of inflation respectively, with a wide GW frequency band, $f_\GW \in [2 \cdot 10^{-10}, 2]$ Hz.

\section{Conclusions}
\label{concl}

This work was devoted to the analysis of PBH formation and the associated production of secondary GWs in the framework of Fibre Inflation which is a class of inflationary models derived within the low energy effective action of a UV complete theory, string theory. This guarantees the freedom to tune the parameters of the inflaton potential since they are related to the microscopic features of the compactification. Moreover, it enlightens the robustness of the model and its stability under quantum corrections due to the underlying symmetries of string theory. 

We first provided appropriate parameter sets allowing for an inflationary regime with an USR period responsible for an enhancement of the scalar power spectrum at different momentum scales larger than CMB ones. These scenarios allow for successful PBH formation with any PBH contribution to DM at different PBH mass scales, extending previous results \cite{Cicoli:2018asa}. Our findings are compatible with current observational bounds also at CMB scales since they lead to a scalar spectral index of order $n_s\simeq 0.965$ and a rather large tensor-to-scalar ratio at the edge of detectability, $r\simeq 0.02$.

These models lead also to a detectable stochastic background of secondary GWs sourced by the enhancement of the scalar power spectrum responsible for PBH formation. Depending on the PBH mass scale which, in turn, results in a different range of GW frequencies, a GW signal could be seen by either LISA, BBO or ET, but  not by higher frequency interferometers like aLIGO-VIRGO since they would correspond to frequencies associated to PBH mass scales $M_\PBH \lesssim 10^{-17}\,M_\odot$ which are strongly constrained by Hawking evaporation. 

Interestingly, we found that secondary GWs could be detectable even if the PBH contribution to the total DM abundance is exponentially suppressed. The only requirement to have a GW energy density large enough to be detected is an enhancement of the scalar power spectrum induced by an USR period between CMB horizon exit and the end of inflation. We therefore realised that the assumption to focus on PBH masses which are not constrained by observations can be relaxed. In fact, since secondary GWs can be observed even for a tiny PBH contribution to DM, we can consider the whole available range of PBH masses from $10^{-17} M_\odot$ to $10^3 M_\odot$ given that the constraints will not affect the peak value of the GW energy density sensibly enough to exit the interferometer sensitivity curves. 

In terms of detectability of secondary GWs by forthcoming interferometers, we therefore conclude that a non-observation would rule out both PBH DM for the whole range $10^{-17} M_\odot \lesssim M_\PBH \lesssim 10^{-12} M_\odot$ where PBHs could still constitute a considerable fraction of the total DM abundance, and an USR epoch in the range $15 \lesssim N \lesssim 30$ (and possibly even higher) efoldings before the end of inflation. As a final remark, we stress that these features are not expected to be exclusive of Fibre Inflation, since they are rather generic for any single-field inflationary model with a potential that has a near inflection point, regardless of its microscopic origin.

\bibliographystyle{JHEP}

\begin{thebibliography}{99}

\bibitem{Hawking:1971ei}
S.~Hawking,
Mon. Not. Roy. Astron. Soc. \textbf{152} (1971), 75

\bibitem{Ivanov:1994pa}
P.~Ivanov, P.~Naselsky and I.~Novikov,
Phys. Rev. D \textbf{50} (1994), 7173-7178
doi:10.1103/PhysRevD.50.7173

\bibitem{Bartolo:2018evs}
N.~Bartolo, V.~De Luca, G.~Franciolini, A.~Lewis, M.~Peloso and A.~Riotto,
Phys. Rev. Lett. \textbf{122} (2019) no.21, 211301
doi:10.1103/PhysRevLett.122.211301
[arXiv:1810.12218 [astro-ph.CO]].

\bibitem{Ananda:2006af}
K.~N.~Ananda, C.~Clarkson and D.~Wands,
Phys. Rev. D \textbf{75} (2007), 123518
doi:10.1103/PhysRevD.75.123518
[arXiv:gr-qc/0612013 [gr-qc]].

\bibitem{Baumann:2007zm}
D.~Baumann, P.~J.~Steinhardt, K.~Takahashi and K.~Ichiki,
Phys. Rev. D \textbf{76} (2007), 084019
doi:10.1103/PhysRevD.76.084019
[arXiv:hep-th/0703290 [hep-th]].

\bibitem{Espinosa:2018eve}
J.~R.~Espinosa, D.~Racco and A.~Riotto,
JCAP \textbf{09} (2018), 012
doi:10.1088/1475-7516/2018/09/012
[arXiv:1804.07732 [hep-ph]].

\bibitem{Kohri:2018awv}
K.~Kohri and T.~Terada,
Phys. Rev. D \textbf{97} (2018) no.12, 123532
doi:10.1103/PhysRevD.97.123532
[arXiv:1804.08577 [gr-qc]].

\bibitem{Domenech:2021ztg}
G.~Dom\`enech,
Universe \textbf{7} (2021) no.11, 398
doi:10.3390/universe7110398
[arXiv:2109.01398 [gr-qc]].

\bibitem{Robson:2018ifk}
T.~Robson, N.~J.~Cornish and C.~Liu,
Class. Quant. Grav. \textbf{36} (2019) no.10, 105011
doi:10.1088/1361-6382/ab1101
[arXiv:1803.01944 [astro-ph.HE]].

\bibitem{Thrane:2013oya}
E.~Thrane and J.~D.~Romano,
Phys. Rev. D \textbf{88} (2013) no.12, 124032
doi:10.1103/PhysRevD.88.124032
[arXiv:1310.5300 [astro-ph.IM]].

\bibitem{Bavera:2021wmw}
S.~S.~Bavera, G.~Franciolini, G.~Cusin, A.~Riotto, M.~Zevin and T.~Fragos,
Astron. Astrophys. \textbf{660} (2022), A26
doi:10.1051/0004-6361/202142208
[arXiv:2109.05836 [astro-ph.CO]].


\bibitem{Maggiore:2019uih}
M.~Maggiore, C.~Van Den Broeck, N.~Bartolo, E.~Belgacem, D.~Bertacca, M.~A.~Bizouard, M.~Branchesi, S.~Clesse, S.~Foffa and J.~Garc\'\i{}a-Bellido, \textit{et al.}
JCAP \textbf{03} (2020), 050
doi:10.1088/1475-7516/2020/03/050
[arXiv:1912.02622 [astro-ph.CO]].




\bibitem{Garcia-Bellido:2017mdw}
J.~Garcia-Bellido and E.~Ruiz Morales,
Phys. Dark Univ. \textbf{18} (2017), 47-54
doi:10.1016/j.dark.2017.09.007
[arXiv:1702.03901 [astro-ph.CO]].

\bibitem{Ezquiaga:2017fvi}
J.~M.~Ezquiaga, J.~Garcia-Bellido and E.~Ruiz Morales,
Phys. Lett. B \textbf{776} (2018), 345-349
doi:10.1016/j.physletb.2017.11.039
[arXiv:1705.04861 [astro-ph.CO]].

\bibitem{Germani:2017bcs}
C.~Germani and T.~Prokopec,
Phys. Dark Univ. \textbf{18} (2017), 6-10
doi:10.1016/j.dark.2017.09.001
[arXiv:1706.04226 [astro-ph.CO]].

\bibitem{Ballesteros:2017fsr}
G.~Ballesteros and M.~Taoso,
Phys. Rev. D \textbf{97} (2018) no.2, 023501
doi:10.1103/PhysRevD.97.023501
[arXiv:1709.05565 [hep-ph]].

\bibitem{Cicoli:2008gp}
M.~Cicoli, C.~P.~Burgess and F.~Quevedo,
JCAP \textbf{03} (2009), 013
doi:10.1088/1475-7516/2009/03/013
[arXiv:0808.0691 [hep-th]].

\bibitem{Broy:2015zba}
B.~J.~Broy, D.~Ciupke, F.~G.~Pedro and A.~Westphal,
JCAP \textbf{01} (2016), 001
doi:10.1088/1475-7516/2016/01/001
[arXiv:1509.00024 [hep-th]].

\bibitem{Cicoli:2016chb}
M.~Cicoli, D.~Ciupke, S.~de Alwis and F.~Muia,
JHEP \textbf{09} (2016), 026
doi:10.1007/JHEP09(2016)026
[arXiv:1607.01395 [hep-th]].

\bibitem{Burgess:2016owb}
C.~P.~Burgess, M.~Cicoli, S.~de Alwis and F.~Quevedo,
JCAP \textbf{05} (2016), 032
doi:10.1088/1475-7516/2016/05/032
[arXiv:1603.06789 [hep-th]].

\bibitem{Burgess:2014tja}
C.~P.~Burgess, M.~Cicoli, F.~Quevedo and M.~Williams,
JCAP \textbf{11} (2014), 045
doi:10.1088/1475-7516/2014/11/045
[arXiv:1404.6236 [hep-th]].

\bibitem{Burgess:2020qsc}
C.~P.~Burgess, M.~Cicoli, D.~Ciupke, S.~Krippendorf and F.~Quevedo,
Fortsch. Phys. \textbf{68} (2020) no.10, 2000076
doi:10.1002/prop.202000076
[arXiv:2006.06694 [hep-th]].

\bibitem{Cicoli:2011it}
M.~Cicoli, M.~Kreuzer and C.~Mayrhofer,
JHEP \textbf{02} (2012), 002
doi:10.1007/JHEP02(2012)002
[arXiv:1107.0383 [hep-th]].

\bibitem{Cicoli:2016xae}
M.~Cicoli, F.~Muia and P.~Shukla,
JHEP \textbf{11} (2016), 182
doi:10.1007/JHEP11(2016)182
[arXiv:1611.04612 [hep-th]].

\bibitem{Cicoli:2017axo}
M.~Cicoli, D.~Ciupke, V.~A.~Diaz, V.~Guidetti, F.~Muia and P.~Shukla,
JHEP \textbf{11} (2017), 207
doi:10.1007/JHEP11(2017)207
[arXiv:1709.01518 [hep-th]].

\bibitem{Starobinsky:1980te}
A.~A.~Starobinsky,
Phys. Lett. B \textbf{91} (1980), 99-102
doi:10.1016/0370-2693(80)90670-X

\bibitem{Kallosh:2013maa}
R.~Kallosh and A.~Linde,
JCAP \textbf{10} (2013), 033
doi:10.1088/1475-7516/2013/10/033
[arXiv:1307.7938 [hep-th]].

\bibitem{Kallosh:2017wku}
R.~Kallosh, A.~Linde, D.~Roest, A.~Westphal and Y.~Yamada,
JHEP \textbf{02} (2018), 117
doi:10.1007/JHEP02(2018)117
[arXiv:1707.05830 [hep-th]].

\bibitem{Cicoli:2018asa}
M.~Cicoli, V.~A.~Diaz and F.~G.~Pedro,
JCAP \textbf{06} (2018), 034
doi:10.1088/1475-7516/2018/06/034
[arXiv:1803.02837 [hep-th]].

\bibitem{Planck:2018vyg}
N.~Aghanim \textit{et al.} [Planck],
Astron. Astrophys. \textbf{641} (2020), A6
[erratum: Astron. Astrophys. \textbf{652} (2021), C4]
doi:10.1051/0004-6361/201833910
[arXiv:1807.06209 [astro-ph.CO]].

\bibitem{Ballesteros:2020qam}
G.~Ballesteros, J.~Rey, M.~Taoso and A.~Urbano,
JCAP \textbf{07} (2020), 025
doi:10.1088/1475-7516/2020/07/025
[arXiv:2001.08220 [astro-ph.CO]].

\bibitem{Cicoli:2020bao}
M.~Cicoli and E.~Di Valentino,
Phys. Rev. D \textbf{102} (2020) no.4, 043521
doi:10.1103/PhysRevD.102.043521
[arXiv:2004.01210 [astro-ph.CO]].

\bibitem{Cicoli:2018tcq}
M.~Cicoli, D.~Ciupke, C.~Mayrhofer and P.~Shukla,
JHEP \textbf{05} (2018), 001
doi:10.1007/JHEP05(2018)001
[arXiv:1801.05434 [hep-th]].

\bibitem{Cicoli:2019ulk}
M.~Cicoli, V.~Guidetti and F.~G.~Pedro,
JCAP \textbf{05} (2019), 046
doi:10.1088/1475-7516/2019/05/046
[arXiv:1903.01497 [hep-th]].

\bibitem{Cicoli:2021yhb}
M.~Cicoli, V.~Guidetti, F.~Muia, F.~G.~Pedro and G.~P.~Vacca,
[arXiv:2107.03391 [astro-ph.CO]].

\bibitem{Cicoli:2021itv}
M.~Cicoli, V.~Guidetti, F.~Muia, F.~G.~Pedro and G.~P.~Vacca,
[arXiv:2107.12392 [hep-th]].

\bibitem{Cicoli:2013oba}
M.~Cicoli, S.~Downes and B.~Dutta,
JCAP \textbf{12} (2013), 007
doi:10.1088/1475-7516/2013/12/007
[arXiv:1309.3412 [hep-th]].

\bibitem{Pedro:2013pba}
F.~G.~Pedro and A.~Westphal,
JHEP \textbf{04} (2014), 034
doi:10.1007/JHEP04(2014)034
[arXiv:1309.3413 [hep-th]].

\bibitem{Cicoli:2014bja}
M.~Cicoli, S.~Downes, B.~Dutta, F.~G.~Pedro and A.~Westphal,
JCAP \textbf{12} (2014), 030
doi:10.1088/1475-7516/2014/12/030
[arXiv:1407.1048 [hep-th]].

\bibitem{Cabella:2017zsa}
P.~Cabella, A.~Di Marco and G.~Pradisi,
Phys. Rev. D \textbf{95} (2017) no.12, 123528
doi:10.1103/PhysRevD.95.123528
[arXiv:1704.03209 [astro-ph.CO]].

\bibitem{Cicoli:2018cgu}
M.~Cicoli and G.~A.~Piovano,
JCAP \textbf{02} (2019), 048
doi:10.1088/1475-7516/2019/02/048
[arXiv:1809.01159 [hep-th]].

\bibitem{Antusch:2017flz}
S.~Antusch, F.~Cefala, S.~Krippendorf, F.~Muia, S.~Orani and F.~Quevedo,
JHEP \textbf{01} (2018), 083
doi:10.1007/JHEP01(2018)083
[arXiv:1708.08922 [hep-th]].

\bibitem{Anguelova:2009ht}
L.~Anguelova, V.~Calo and M.~Cicoli,
JCAP \textbf{10} (2009), 025
doi:10.1088/1475-7516/2009/10/025
[arXiv:0904.0051 [hep-th]].

\bibitem{Berg:2005ja}
M.~Berg, M.~Haack and B.~Kors,
JHEP \textbf{11} (2005), 030
doi:10.1088/1126-6708/2005/11/030
[arXiv:hep-th/0508043 [hep-th]].

\bibitem{Berg:2007wt}
M.~Berg, M.~Haack and E.~Pajer,
JHEP \textbf{09} (2007), 031
doi:10.1088/1126-6708/2007/09/031
[arXiv:0704.0737 [hep-th]].

\bibitem{Cicoli:2007xp}
M.~Cicoli, J.~P.~Conlon and F.~Quevedo,
JHEP \textbf{01} (2008), 052
doi:10.1088/1126-6708/2008/01/052
[arXiv:0708.1873 [hep-th]].

\bibitem{Kachru:2003aw}
S.~Kachru, R.~Kallosh, A.~D.~Linde and S.~P.~Trivedi,
Phys. Rev. D \textbf{68} (2003), 046005
doi:10.1103/PhysRevD.68.046005
[arXiv:hep-th/0301240 [hep-th]].

\bibitem{Cicoli:2015ylx}
M.~Cicoli, F.~Quevedo and R.~Valandro,
JHEP \textbf{03} (2016), 141
doi:10.1007/JHEP03(2016)141
[arXiv:1512.04558 [hep-th]].

\bibitem{Cicoli:2012fh}
M.~Cicoli, A.~Maharana, F.~Quevedo and C.~P.~Burgess,
JHEP \textbf{06} (2012), 011
doi:10.1007/JHEP06(2012)011
[arXiv:1203.1750 [hep-th]].

\bibitem{Gallego:2017dvd}
D.~Gallego, M.~C.~D.~Marsh, B.~Vercnocke and T.~Wrase,
JHEP \textbf{10} (2017), 193
doi:10.1007/JHEP10(2017)193
[arXiv:1707.01095 [hep-th]].

\bibitem{Motohashi:2017kbs}
H.~Motohashi and W.~Hu,
Phys. Rev. D \textbf{96} (2017) no.6, 063503
doi:10.1103/PhysRevD.96.063503
[arXiv:1706.06784 [astro-ph.CO]].

\bibitem{Inomata:2017okj}
K.~Inomata, M.~Kawasaki, K.~Mukaida, Y.~Tada and T.~T.~Yanagida,
Phys. Rev. D \textbf{96} (2017) no.4, 043504
doi:10.1103/PhysRevD.96.043504
[arXiv:1701.02544 [astro-ph.CO]].

\bibitem{Atal:2018neu}
V.~Atal and C.~Germani,
Phys. Dark Univ. \textbf{24} (2019), 100275
doi:10.1016/j.dark.2019.100275
[arXiv:1811.07857 [astro-ph.CO]].

\bibitem{Atal:2021jyo}
V.~Atal and G.~Dom\`enech,
JCAP \textbf{06} (2021), 001
doi:10.1088/1475-7516/2021/06/001
[arXiv:2103.01056 [astro-ph.CO]].

\bibitem{Atal:2019erb}
V.~Atal, J.~Cid, A.~Escriv\`a and J.~Garriga,
JCAP \textbf{05} (2020), 022
doi:10.1088/1475-7516/2020/05/022
[arXiv:1908.11357 [astro-ph.CO]].

\bibitem{Germani:2018jgr}
C.~Germani and I.~Musco,
Phys. Rev. Lett. \textbf{122} (2019) no.14, 141302
doi:10.1103/PhysRevLett.122.141302
[arXiv:1805.04087 [astro-ph.CO]].

\bibitem{Lyth:1996im}
D.~H.~Lyth,
Phys. Rev. Lett. \textbf{78} (1997), 1861-1863
doi:10.1103/PhysRevLett.78.1861
[arXiv:hep-ph/9606387 [hep-ph]].

\bibitem{Acquaviva:2002ud}
V.~Acquaviva, N.~Bartolo, S.~Matarrese and A.~Riotto,
Nucl. Phys. B \textbf{667} (2003), 119-148
doi:10.1016/S0550-3213(03)00550-9
[arXiv:astro-ph/0209156 [astro-ph]].

\bibitem{Maggiore:1999vm}
M.~Maggiore,
Phys. Rept. \textbf{331} (2000), 283-367
doi:10.1016/S0370-1573(99)00102-7
[arXiv:gr-qc/9909001 [gr-qc]].

\bibitem{Bartolo:2018rku}
N.~Bartolo, V.~De Luca, G.~Franciolini, M.~Peloso, D.~Racco and A.~Riotto,
Phys. Rev. D \textbf{99} (2019) no.10, 103521
doi:10.1103/PhysRevD.99.103521
[arXiv:1810.12224 [astro-ph.CO]].

\end{thebibliography}

\end{document}